\documentclass[twocolumn,showpacs,preprintnumbers,amsmath,amssymb]{revtex4}
\usepackage[dvips]{graphicx}
\usepackage{dcolumn}

\begin{document}

\title
{Double-Core Vortex Stabilized by Disorder in Superfluid $^3$He B Phase in Globally Isotropic Aerogel}

\author{Natsuo Nagamura and Ryusuke Ikeda}

\affiliation{%
Department of Physics, Kyoto University, Kyoto 606-8502, Japan
}

\date{\today}

\begin{abstract} 
@In a $p$-wave Fermi superfluid suffering from the nonmagnetic impurity scatterings, a coefficient of a gradient term becomes divergent upon cooling. Consequences of this divergent rigidity in the stable vortices in the B phase in {\it globally isotropic} aerogel are considered where the "impurity scattering" events are brought by the aerogel structure. For a moderately strong "impurity scatterings", the superfluid transition line $T_c(P)$ has a quantum critical point at a low but finite pressure. We find that, with decreasing $T_c$ via the lowering of the pressure, the distance between the half cores composing the core of the nonaxisymmetric double-core vortex which is stable at lower pressures grows as a result of the rigidity diverging at lowering temperature. The obtained result is compared with the elongation of the half core pair arising from the Fermi-liquid correction. 
\end{abstract}


%\keyword{}

\maketitle

\section{Inroduction}
\label{sec:intro}
 Study of superfluid $^3$He in aerogel has begun with much interest in how the pairing state is affected by the "impurity scattering" events stemming from the structure of aerogel which is a porous material \cite{Review}. It has been clarified that, when the scattering events due to the aerogel has a global anisotropy, pairing states to be realized may be different from those in the pure bulk liquid \cite{AI06}. In contrast, in the case of liquid $^3$He in aerogels with no global anisotropy, it is believed at present that the only pairing state to be realized in equilibrium is the B phase which is the ground state in the weak coupling approximation valid at lower pressures \cite{Review}. Although the pairing state to be realized is conventional, it is unclear whether the resulting excitation remains conventional and unaffected by the impurity scatterings. 

It is well known that the impurity scattering is relevant to the $p$-wave paired Fermi superfluid state and diminishes the superfluid transition temperature $T_c$. In the case of the liquid $^3$He, the resulting pressure-dependent transition curve $T_c(P)$ has a quantum critical pressure $P_c$ ($> 0$) \cite{Matsumoto}. The resulting gradient term of the Ginzburg-Landau (GL) free energy close to $P_c$ is found to have a $|{\rm log}T|$-divergent coefficient \cite{AdaJPSJ}. This $|{\rm log}T|$-dependence in the gradient term is a consequence of the cancellation between the relaxation rate due to the impurity scattering and the impurity-induced vertex correction and has the same origin as that in the mass term in the $s$-wave pairing case leading to the absence of any impurity effect on $T_c$ there \cite{Anderson}. It will be valuable to examine a consequence of this divergent rigidity in the vortex solution in B-phase. 

In this work, the stable vortex solution in the superfluid $^3$He B phase under the "impurity scattering" modelling the system in the globally isotropic aerogel is examined at low pressures. Since we focus on the region close to $T_c(P)$ at lower pressures, throughout this work we use the weak-coupling approximation in which the stable vortex is the nonaxisymmetric or the double-core one. Performing the direct 2D numerical computation, the temperature dependence of the core structure of the double-core vortex will be examined in the Ginzburg-Landau (GL) approach \cite{Th}. It is found that the rigidity growing upon cooling enhances the distance between the half cores composing the double-core vortex and thus. makes the vortex core more anisotropic. It has been pointed out in the quasiclassical approach \cite{Th2} that the Fermi-liquid correction neglected in the conventional GL approach will also enhance the vortex core anisotropy. The Fermi-liquid correction to the gradient term formulated in the GL approach elsewhere \cite{Nagamura} is also found to enhance the size of the half core pair. Therefore, the present disorder-induced mechanism which stabilizes the double-core vortex but is effective even close to $T_c$ is in essence different from that due to the Fermi-liquid correction. 

This paper is organized as follows. In sec.2, the model and the details of our analysis are explained, and the obtained numerical results are discussed in sec.3. The content of the present work is summarized in sec.4. 

\section{Model}

The starting model Hamiltonian of our analysis in this work is the mean field BCS Hamiltonian with the momentum-independent impurity scattering term 
\begin{equation}
{\cal H}_{\rm imp} = \int d^3{\bf r} \sum_\sigma {\hat \psi}^\dagger_\sigma({\bf r}) \, u({\bf r}) \, {\hat \psi}_\sigma({\bf r}) 
\end{equation} 
with ${\overline {u({\bf r})}}=0$ and 
\begin{equation}
{\overline {|u_{\bf k}|^2}} = \frac{1}{2 \pi N(0) \tau_o}, 
\label{impline}
\end{equation}
where the overbar denotes the random average, $N(0)$ is the density of states per spin on the Fermi surface in the normal state, $\tau_o$ is the relaxation time, and ${\hat \psi}$ is the fermion operator. The quasiparticle Green's function takes the form 
\begin{equation}
G_\varepsilon({\bf p}) = \frac{1}{{\rm i} \, {\tilde \varepsilon} - \xi_{\bf p}}, 
\end{equation}
where 
${\tilde \varepsilon} = \varepsilon + {\rm sgn}(\varepsilon)/(2 \tau_o)$, and $\varepsilon$ is a fermionic Matsubara frequency, and ${\rm sgn}(\varepsilon) = \varepsilon/|\varepsilon|$. 
Further, we need to incorporate the impurity scattering-induced correction to the pairing vertex (see Fig.1) to construct the gradient terms of the GL free energy which we use as a tool to examine a single-vortex solution. In the present case where the scattering event is isotropic, the pairing vertex function $\Gamma_i(\varepsilon, {\hat {\bf p}}, {\bf q})$ satisfying the relation 
\begin{figure}[tbp]
\begin{center}
{
\includegraphics[scale = 0.5]{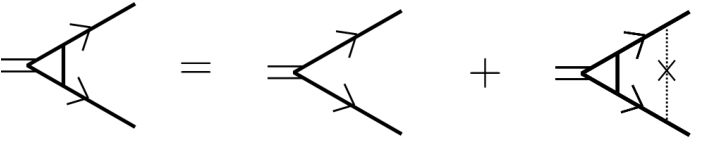}
}
\caption{Diagrammatic representation of the impurity-scattering induced vertex correction to the pairing process. Each solid line with an arrow denotes the quasiparticle Green's function, the double line denotes the pair-field, and the thin line with a cross denotes the impurity scattering strength occurring after the random-average. 
}
\label{s:fig:VC}
\end{center}
\end{figure}
\begin{eqnarray}
\Gamma_j(\varepsilon, {\hat {\bf p}}, {\bf q}) &=& {\hat {\bf p}}_j + \frac{1}{2 \pi N(0) \tau_o} \, \int_{\bf {p'}} \Gamma_j(\varepsilon, {\hat {\bf {p'}}}, {\bf q}) \nonumber \\
&\times& G_\varepsilon({\bf {p}'}+{\bf q}/2) \, G_{-\varepsilon}(-{\bf {p}'}+{\bf q}/2) 
\end{eqnarray}
%\end{widetext}
according to Fig.1 has the solution of the form 
\begin{equation}
\Gamma_j = {\hat p}_j - {\rm i} {\rm sgn}(\varepsilon) v_{\rm F} q_j \, B_0
\end{equation} 
up to the lowest order in the external wavevector ${\bf q}$, where $v_{\rm F}$ is the Fermi velocity, and ${\hat {\bf p}}$ is the unit vector of ${\bf p}$. By carrying out the ${\bf {p}}'$-integral, the second term of r.h.s. of eq.(4) becomes 
\begin{equation}
\frac{1}{2|{\tilde \varepsilon}|\tau_o} 
\biggl\langle \Gamma_j(\varepsilon, {\hat {\bf {p}}}', {\bf q}) \biggl(1 - {\rm i} {\rm sgn}(\varepsilon) v_{\rm F} \frac{{\hat {p}'}\cdot{\bf q}}{2|{\tilde \varepsilon}|} \biggr) \biggr\rangle. 
\end{equation}
where $\langle \,\,\,\ \rangle$ denotes the angle average over the Fermi surface. 
Then, $B_0$ becomes 
\begin{equation}
B_0 = \frac{\langle {\tilde p}_z^2 \rangle}{4 |{\tilde \varepsilon} \varepsilon| \tau_o} = \frac{1}{6 |\varepsilon| (1 + 2 \tau_o |\varepsilon|)}.
\end{equation}
The behavior $B_0 \propto |\varepsilon|^{-1}$ implies cancellation between the selfenergy and the vertex correction due to the "impurity scattering". In the conventional $s$-wave paired superconductor, a similar cancellation occurs in the quadratic mass term of the GL free energy and leads to the celebrated argument of the impurity-independence of the mean field superconducting transition temperature \cite{Anderson}. 
It was previously argued \cite{AdaJPSJ} that this cancellation in the gradient term affects the fluctuation conductivity close to a quantum critical point of a $p$-wave paired superconducting transition. 

Using the vertex correction derived above, the quadratic terms of the GL free energy to be used in our numerical work are obtained through the expression 
\begin{widetext}
\begin{equation}
F_{2} = \sum_{\bf q} \biggl[\frac{N(0)}{3} \biggl({\rm ln}\biggl(\frac{T}{T_{c0}}\biggr) + T \sum_\varepsilon \frac{\pi}{|\varepsilon|} \biggr) \delta_{i,j} - T \sum_\varepsilon \int_{\bf p} {\hat p}_i \Gamma_j(\varepsilon, {\bf p}, {\bf q}) G_\varepsilon({\bf p}+{\bf q}/2) G_{-\varepsilon}(-{\bf p}+{\bf q}/2) \biggr] A^*_{\mu,i}({\bf q}) A_{\mu,j}({\bf q}). 
\end{equation}
\end{widetext}
Performing the ${\bf p}$-integral, the resulting weak-coupling GL free energy including the quartic terms takes the form 
\begin{eqnarray}
F &=& \int_{\bf r} (f_2 + f_4) \nonumber\\
f_4 &=& \beta_1|A_{\mu i}A_{\mu i}|^2+\beta_2(A_{\mu i}A_{\mu i}^*)^2 + \beta_3 A_{\mu i}^*A_{\nu i}^*A_{\mu j}A_{\nu j} \nonumber \\ 
&+& \beta_4 A_{\mu i}^*A_{\nu i}A_{\nu j}^*A_{\mu j} + \beta_5 A_{\mu i}^*A_{\nu i}A_{\nu j}A_{\mu j}^*, \nonumber\\
f_2 &=& \alpha A_{\mu i}A^*_{\mu i} + 2 K_1 \partial_i A_{\mu i}\partial_j A_{\mu j}^* + K_2 \partial_i A_{\mu j}\partial_i A_{\mu j}^*. 
\end{eqnarray}
Here, the coefficients $K_n$ are expressed as 
\begin{eqnarray}
K_1 &=& K_2 - \frac{1}{36} \biggl( \frac{v_{\rm F}}{2 \pi T} \biggr)^2 \biggl[8(2 \pi \tau T)^2 \biggl(\psi^{(0)}(1/2) \nonumber \\
&-& \psi^{(0)}(y) \biggr) + 4(2 \pi \tau T) \psi^{(1)}(y) - \psi^{(2)}(y) \biggr] \nonumber \\
K_2 &=& - \frac{1}{120} \biggl(\frac{v_{\rm F}}{2 \pi T} \biggr)^2 \psi^{(2)}(y), \nonumber\\
\end{eqnarray} 

%from vortex
in the representation of the center-of-mass coordinate of Cooper-pairs, where $y=1/2 + 1/(4 \pi \tau_0 T)$, and $\psi^{(n)}(x)$ denotes the $n$-times derivative of the di-gamma function $\psi^{(0)}(x) = {\rm const} - \sum_{n \geq 0}(n+x)^{-1}$. The presence of the $\psi^{(0)}(1/2)-\psi^{(0)}(y)$ term in the expression of $K_1$ leads to the $|{\rm log}T|$-divergence of this coefficient. On the other hand, regarding $\alpha$ and the coefficients in $F_4$, the same expressions as those in Ref.\cite{AI06} will be used by setting the anisotropy parameter $\delta_u$ to be zero. 

Strictly speaking, the $|{\rm log}T|$-divergence of $K_1$ should be cut off by higher order spatial-gradients or the wavenumber $\sim (|\alpha(T=0)|/K_2)^{1/2}$ measuring the distance from the quantum critical pressure at $T=0$. Note that the present GL analysis is originally justified at least close to the $T_c(P)$-curve. Thus, it is possible that some of our results taken at low enough temperatures will not be quantitatively sufficient. 

\section{Numerical Results}
\label{s:sec:B}
To study the stability of the double-core vortex structure of the B phase in the disordered media, we have numerically solved the variational equations for the GL free energy, eq.(9). As the numerical method, we have followed the two-dimensional analysis developed by Thuneberg \cite{Th,Nagamura}. To make the convergence of computation better, the "London" representation \cite{VolovikB} of the double-core structure will be used as the initial condition of our numerical analysis: 
\begin{equation}
A_{\mu,j} = \exp({\rm i}\Phi(a)) [ {\hat x}_\mu {\hat m}_j + {\hat y}_\mu {\hat n}_j + {\hat z}_\mu ({\hat m} \times {\hat n})_j],  
\end{equation}
where $\Phi(a) = (\phi_+ + \phi_-)/2$, $\phi_\pm = {\rm tan}^{-1}(y/(x \mp a))$, ${\hat m}_j = {\hat x}_j {\rm cos}[(\phi_+ - \phi_-)/2] - {\hat y}_j {\rm sin}[(\phi_+ - \phi_-)/2]$, ${\hat n}_j = ({\hat z} \times {\hat m})_j$, and ${\hat z}$ is the direction parallel to the vortex lines. Figure 2 indicates the implication of the phase variables $\phi_\pm$. 
\begin{figure}[tbp]
\begin{center}
{
\includegraphics[scale = 0.7]{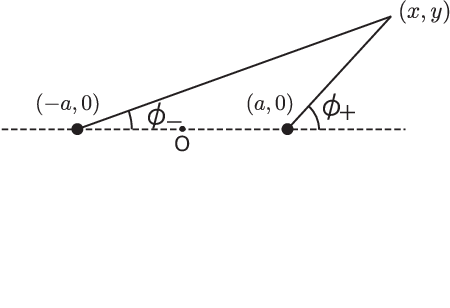}
}
\caption{Schematic figure expressing the structure in the $x$-$y$ plane of a half-core pair expressed in the "London limit". The angle variables $\phi_\pm$ used in eq.(11) are indicated. 
}
\label{s:fig:angles}
\end{center}
\end{figure}

Equation (11) may be rewritten in the form 
\begin{eqnarray}
A_{\mu,j} &=& \exp({\rm i}\phi_+) ({\hat e}_-)_\mu ({\hat e}_+)_j + \exp({\rm i}\phi_-) ({\hat e}_+)_\mu ({\hat e}_-)_j \nonumber \\
&+& \exp({\rm i}\Phi(a)) {\hat z}_\mu {\hat z}_j, 
\end{eqnarray}
which can be regarded as an extension of a HQV pair represented in the chiral basis in the A phase \cite{Ivanov,Nagamura}, where ${\hat e}_\pm = ({\hat x} 
\pm {\hat i} {\hat y})/\sqrt{2}$. 
By assuming a two-fold mirror symmetry of the expected structure of the vortex with its center at the origin $x=y=0$, we have performed the numerical analysis only in $x > 0$ and $y > 0$. Further, the energy $F$, each component of the order parameter $A_{\mu,j}$, and the coordinate ${\bf r}$ will be represented hereafter in the scale transformation 
\begin{eqnarray}
\tilde{A}&=&\frac{A}{\Delta_B}\nonumber\\
\tilde{\mathbf{x}}&=&\frac{\mathbf{x}}{\xi}\nonumber\\
\tilde{F}&=&\frac{F}{6\beta_{12}+2\beta_{345}}, 
\end{eqnarray}
%\begin{eqnarray}
%\frac{\delta \tilde{F}[\tilde{A}(\tilde{\mathbf{x}}),\tilde{A}^*(\tilde{\mathbf%{x}})]}{\delta \tilde{A}^*(\tilde{\mathbf{x}})}=0\nonumber\\
%\end{eqnarray}
where $\Delta_\mathrm{B}=[|\alpha|/(12\beta_{12}+4\beta_{345})]^{1/2}$ is the amplitude of the orderparameter in the bulk B phase, and $\xi=(K_2/|\alpha|)^{1/2}$ is the corresponding coherence length defined based on the fact that $K_1=K_2$ in the bulk B phase in the weak-coupling approach. Further, the system size used in our numerical analysis is $60 \times 60$ in the unit mentioned above, and the grid size is $0.5$ in the same unit. 

By numerically solving the variational equations, the resulting $\tilde{A}$ will be substituted into $\tilde{F}$, and the two-dimensional spatial integrals will be performed over the system size $L$ under the assumption of a straight vortex line extending along the remaining direction, $z$-axis. The resulting vortex energy consists of the two $L$-dependent terms, the bulk energy term $\propto L$ and the ${\rm log}L$-contribution arising from the gradient term, and a constant term sensitive to the details of the vortex core. Since we focus on an energy difference between two vortices, the numerical results will be presented in a way subtracting the two $L$-dependent terms. 
\begin{figure}[tbp]
\begin{center}
{
\includegraphics[scale = 0.4]{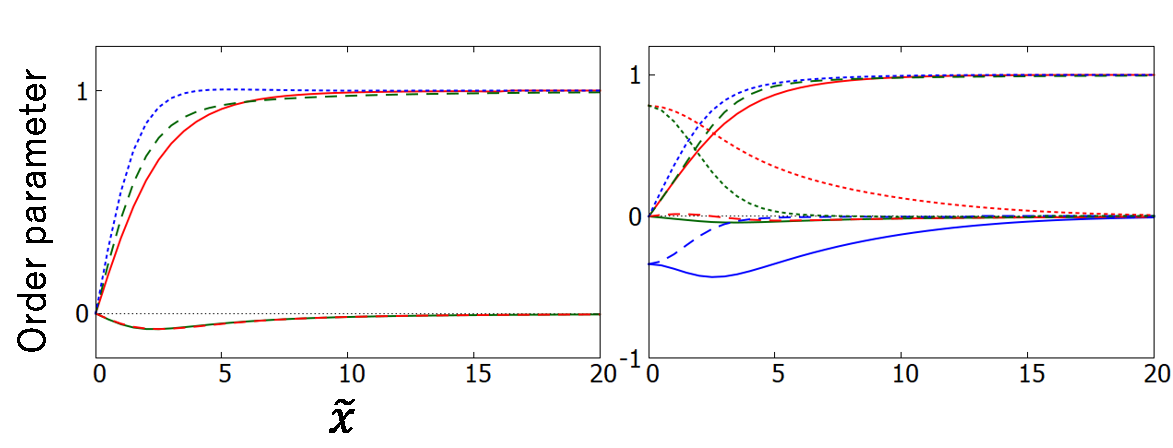}
}
{
\includegraphics[scale = 0.4]{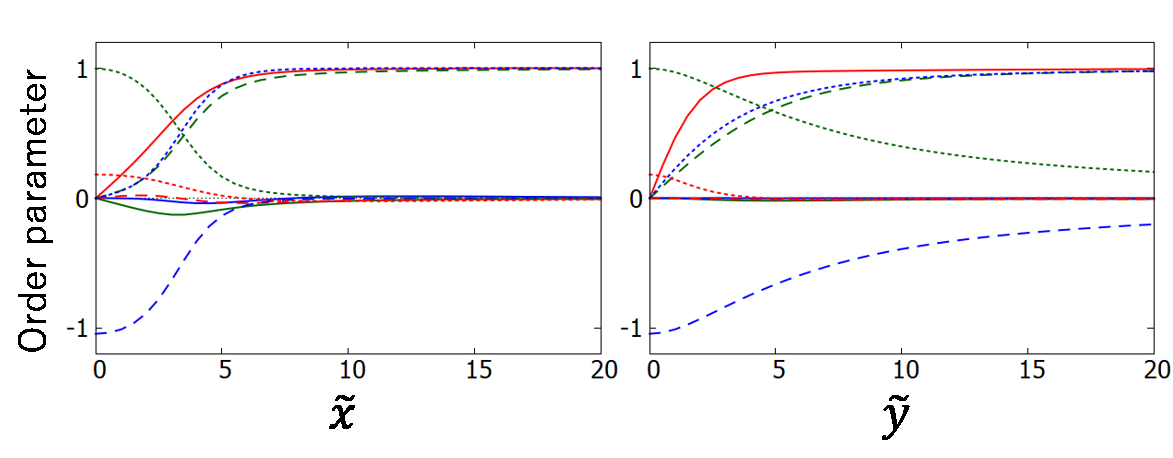}
}
\caption{
Order parameter components $A_{\mu,j}$ (or $-{\rm i} A_{\mu,j}$) close to the center of the o-vortex(upper-left), the v-vortex (upper-right), and the double-core vortex (lower two figures) in the bulk superfluid $^3$He. In the double-core vortex, the two half cores separated along the $x$-axis are present, and the lower left (lower right) figure expresses the $x$-dependence at $y=0$ ($y$-dependence at $x=0$) of the order parameter components. The spin index $\mu$ of the order parameter is classified by the line type, while the orbital index $j$ is distinguished by the color. That is, $\mu=x$ (solid line), $=y$ (dashed curve), and $=z$ (dotted curve), while $j=x$ (red), $=y$ (green), and $=z$ (blue). Further, in the figures, $A_{xz}$, $A_{zx}$, and all of the diagonal components $A_{s,s}$ are real, while other components are purely imaginary and represented as $-{\rm i} A_{\mu,j}$. These prescriptions for describing $A_{\mu,j}$ components are commonly used in all figures hereafter. 
}
\label{s:fig:pureovd}
\end{center}
\end{figure}

\begin{figure}[t]
\begin{center}
{
\includegraphics[scale = 0.5]{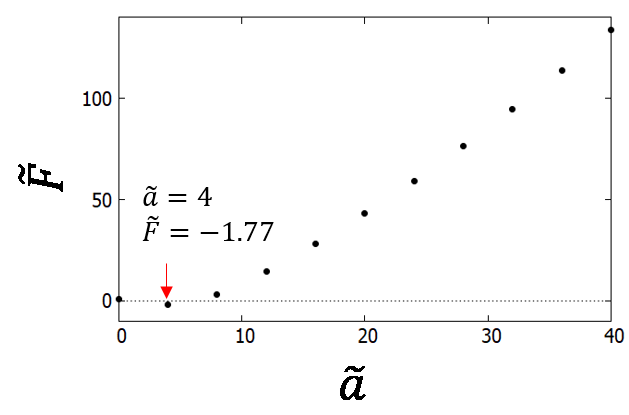}
}
\caption{
The $a$-dependence of the normalized vortex energy $\tilde{F}(\tilde{a})=\frac{F(\tilde{a})-F_v}{F_o-F_v}$ in the case of the bulk liquid, where $F_v$ ($F_o$) is the vortex energy of the v-vortex (o-vortex) given in the upper right (upper left) figure of Fig.3, and $F({\tilde a})$ is the corresponding one of the double-core vortex with the normalized distance between the half cores ${\tilde a}$. 
}
\label{s:fig:pureenergy}
\end{center}
\end{figure}

%\begin{widetext}
\begin{figure}[tbp]
\begin{center}
{
\includegraphics[scale = 0.3]{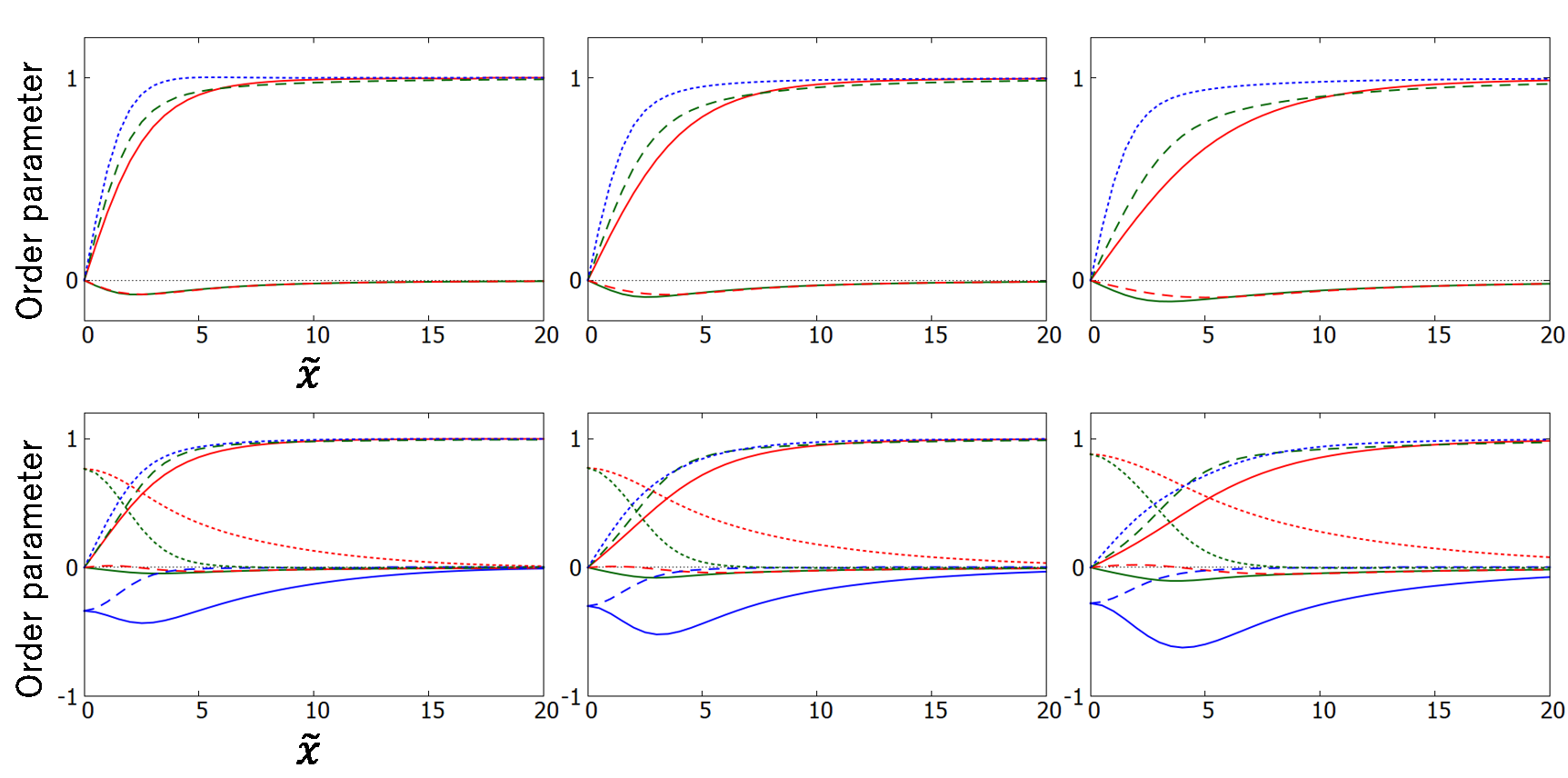}
}
\caption{Order parameters' spatial variations of the o-vortex (upper) and v-vortex (lower) near each core at temperatures $T=1({\rm mK})$ (left), $10 (\mu{\rm K})$ (middle), $0.1(\mu{\rm K})$ (right) on the $T_c(P)$ curve in the impure case. The parameter $\tau$ is fixed to the value $(2 \pi \tau)^{-1} = 0.137 ({\rm mK})$. Definitions of the order parameter components are the same as in Fig.3. 
}
\label{s:fig:Bvortex2}
\end{center}
\end{figure}
%\end{widetext}
%

First, the so-called o-vortex with the normal core, the v-vortex with the A-phase and $\beta$ phase components in its core, and the nonaxisymmetric or the double-core vortex in the bulk B phase will be discussed which were obtained by using the GL free energy, eq. (9), with $\tau^{-1}=0$. In the case of the bulk B phase and when taken in the unit of eq.(13), the temperature does not become a parameter specifying the vortex structure. The resulting core structures of the three types of vortices are shown in Fig.3 and essentially coincide with the corresponding results in previous works \cite{Th}. In fact, as Fig.4 shows, the double-core vortex consisting of a half core pair with the separation ${\tilde a}=4$ becomes the structure with the lowest energy in the weak-coupling approximation. Here, ${\tilde F}$ denotes $\tilde{F}(\tilde{a})=\frac{F(\tilde{a})-F_v}{F_o-F_v}$ expressed in terms of the energy values $F_o,F_v,F(\tilde{a})$ of the o-vortex, the v-vortex, and the double-core one with the half-core separation ${\tilde a}$. The double-core vortex with $a=0$ is essentially the same as the o-vortex which is never minimized 
in energy. Further, the v-vortex always has a lower energy than that of the o-vortex. Thus, a negative ${\tilde F}$ at some $a$ indicates that the double-core vortex has the lowest energy at the same $a$. 
%
%\begin{widetext}
\begin{figure}[tbp]
\begin{center}
{
\includegraphics[scale = 0.3]{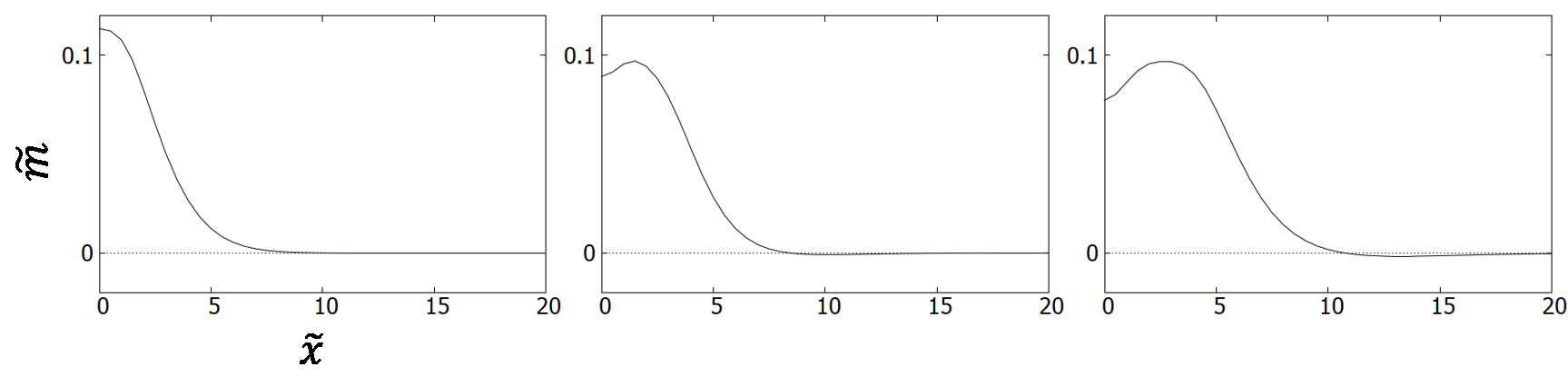}
}
\caption{
Spatial variations of the magnetization arising from the $\beta$-phase components $-{\rm i}A_{xz}$ and $A_{yz}$ shown in the lower figures in Fig.5, expressing  the v-vortex in the impure case. 
}
\label{s:fig:Bvortex}
\end{center}
\end{figure}
%\end{widetext}
%

%\begin{widetext}
\begin{figure}[tbp]
\begin{center}
{
\includegraphics[scale = 0.3]{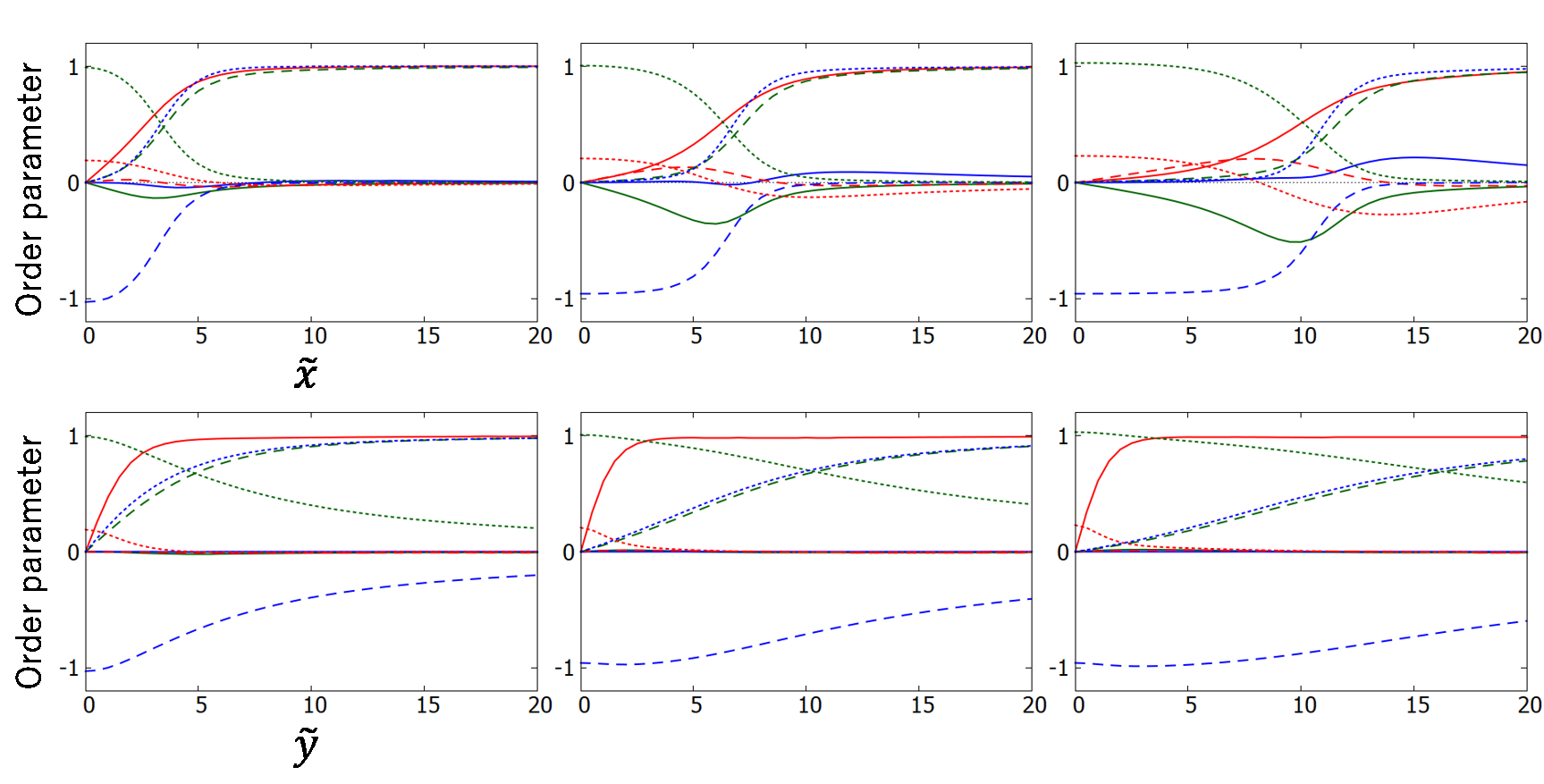}
}
\caption{Order parameters' spatial variations of the double-core vortex at the temperatures $T=1({\rm mK})$ (left), $10(\mu{\rm K})$ (middle), and $0.1(\mu{\rm K})$ (right) in the impure case corresponding to the lower figures of Fig.3 in the pure case. The center of the vortex core is at the origin ($0$, $0$). The order parameter components are defined in the same manner as in Fig.3. The upper figures express their variations parallel to the $x$-direction along which the half core pair is connected by a planar string, while the lower ones express their vertical variations parallel to the $y$-axis. The optimal ${\tilde a}$-value of each figure is found from Figs.8 and 9. 
}
\label{s:fig:double-corevortex}
\end{center}
\end{figure}

%
%\begin{widetext}
\begin{figure}[tbp]
\begin{center}
{
\includegraphics[scale = 0.17]{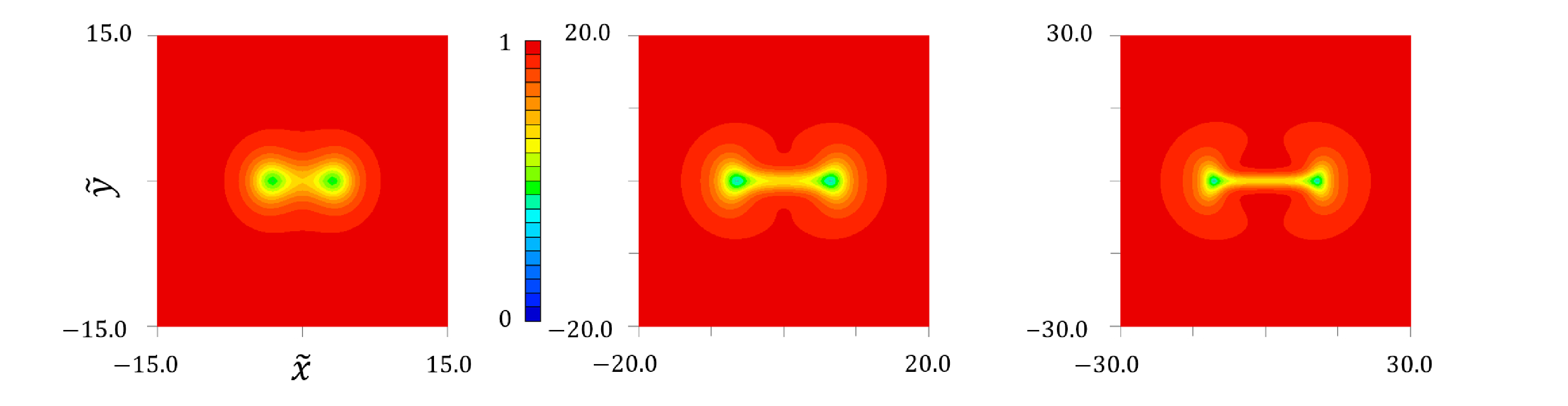}
}
\caption{
Variations of the averaged amplitude $\sum_{\mu,j} |A_{\mu,j}|^2$ in the 2D ($x$-$y$) plane near the double-core vortex at the three temperatures $T=1({\rm mK})$ (left), $10(\mu{\rm K})$ (middle), and $0.1(\mu{\rm K})$ (right). The ${\tilde a}$ value grows with decreasing temperature. 
}
\label{s:fig:rho_B}
\end{center}
\end{figure}
%\end{widetext}

As seen in Fig.4, the energy monotonously increases with increasing $a$. To understand this feature, let us closely examine the spatial variations of the order parameter components in the lower figures of Fig.3. These two figures show that each of the diagonal components $A_{j,j}$ approaches unity far from the vortex center but vanishes close to the origin $x=y=0$, and that $A_{zy} = -A_{yz}=1$ close to the origin. Roughly speaking, this feature can be regarded as appearance of a planar phase's string close to the center and coincides with the representation (11). Therefore, the monotonic increase of the energy with $a$ in Fig.4 is interpreted as a reflection of the tension of the planar string increasing with $a$ which occurs through a cost of the condensation energy.

%
%\begin{widetext}
\begin{figure}[tbp]
\begin{center}
{
\includegraphics[scale = 0.3]{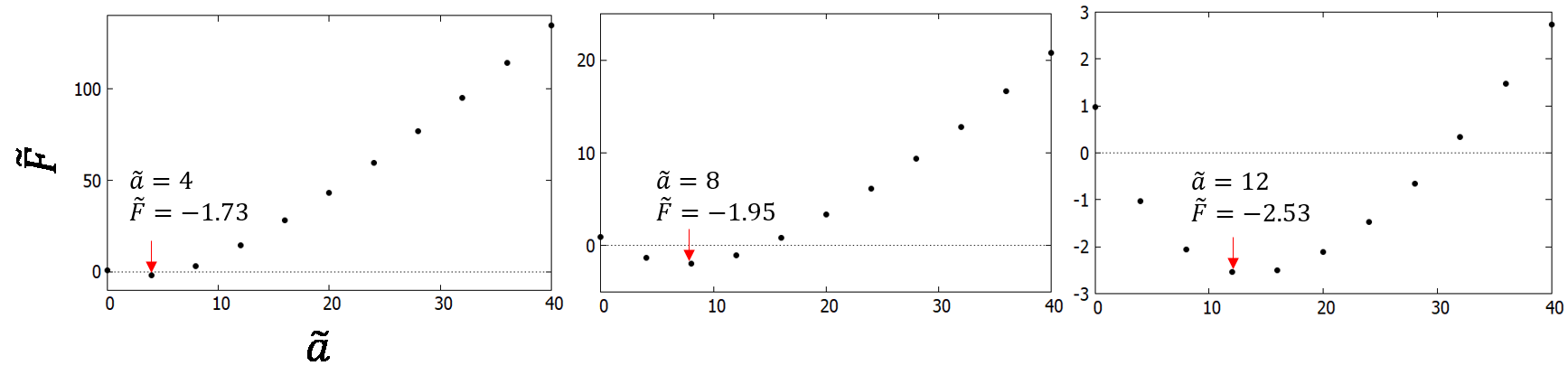}
}
\caption{Normalized energy curves 
${\tilde F}({\tilde a})$ v.s. ${\tilde a}$ of the double-core vortices at $T=1 ({\rm mK})$ (left),$10 (\mu{\rm K})$ (middle), and $0.1(\mu{\rm K})$ (right). 
}
\label{s:fig:energy_B}
\end{center}
\end{figure}
%\end{widetext}

At this time, the same analysis has been performed in the impure ($\tau^{-1} > 0$) case corresponding to $^3$He in a globally isotropic aerogel. Qualitatively, the results on the relative stability of the three vortices, the o-vortex, the v-vortex, and the double-core one, remain unchanged. For instance, the v-vortex is more stable than the o-vortex, and the most stable vortex becomes the double-core vortex with some $a$-value. First, let us briefly explain how the core structure of the v-vortex to be realized at higher pressures is changed by the impurity scatterings with decreasing the temperature. The results corresponding to the upper figures in Fig.3, that is the structures of the o-vortex and v-vortex in the B-phase in an isotropic aerogel, are shown in Fig.5. As the temperature is lowered, the $\beta$-phase component of the order parameter is reduced at the core and becomes maximum rather at a position shifted from the center. As presented in Fig.6, this effect of the impurity scattering is more clearly seen in the spatial distribution of the magnetization accompanying the $\beta$-phase component close to the vortex core. That is, at the vortex center, the spatial distribution of the magnetization shows not a maximum but a local minimum as a result of the impurity-scattering induced reduction of the $\beta$-phase components as the temperature is significantly 
lowered. 

Next, we will turn to the double-core vortex and will examine how the core structure of this vortex is changed when the temperature is lowered along the $T_c(P)$-line. Hereafter, the parameter $\tau$ is fixed to the value $(2 \pi \tau)^{-1} = 0.137 ({\rm mK})$. The order parameter configurations at $T= 1({\rm mK})$, $10(\mu{\rm K})$, and $0.1(\mu{\rm K})$ and at the ${\hat a}$ values minimizing the energy are shown in Fig.7. As Figs.8 and 9 show, the resulting ${\tilde a}$-value minimizing the energy at each temperature remarkably increases with decreasing the temperature along the $T_c(P)$-curve. 

To explain the increase of the size of a half-core pair, the order parameter close to the two half cores will be expressed, following Ref.\cite{VolovikB}, in the "London" representation defined by eq.(11) 
Then, by substituting this expression into the two gradient terms, the energy gain of a half-core pair relative to that of the o-vortex expressed by $A_{\mu,j} = \exp({\rm i}\Phi(0)) [{\hat x}_\mu {\hat x}_j + {\hat y}_\mu {\hat y}_j +{\hat z}_\mu {\hat z}_j]$ becomes 
\begin{equation}
\Delta F(a) = - \frac{\pi}{2} (2 K_1 + K_2) {\rm ln}\biggl(\frac{2a}{\xi_c} \biggr), 
\end{equation}
where $\xi_c$ is the core size of each half core. The optimal $a$-value is determined by balancing $\Delta F(a)$ with the energy of a planar string connecting the two half-cores with each other which, as already mentioned, is proportional to $a$ and is determined by the amplitude change of $A_{z,z}$. Thus, the optimal $a$ is linearly proportional to $K_1$. 
In fact, as the lower three figures of Fig.7 show, the $y$-value over which $A_{zz}$ vanishes is not sensitive to the cooling. It implies that the formation of the planar string is determined by the bulk terms of the free energy unrelated to the coefficient $K_1$ so that the above estimation ${\tilde a} \propto K_1$ will be justified. 

The impurity-induced enhancement of the anisotropy of the core structure of the double-core vortex clarified in Figs.8 and 9 should be distinguished from the increase of the anisotropy induced by the correlation \cite{Th2}. To stress this point, let us briefly explain how the Fermi-liquid (FL) correction term derived in Ref.\cite{Nagamura} within the GL framework leads to an increase of the separation between the two half cores. The FL corrections occurring through the self energy term of the quasiparticle Green's function may be assumed to have already been incorporated in the weak-coupling contribution. In Ref.\cite{Nagamura}, it has been clarified how the remaining FL correction to occur only in the gradient terms appears via a simple extention of the microscopic derivation of the GL functional. Here, we focus on a rough estimation of the resulting FL-corrected contribution to eq.(14). It can be accomplished by applying the "London" representation (11) of the order parameter to eq.(A17) in Ref.\cite{Nagamura}. Then, as the term to be added to eq.(14), we obtain 
\begin{equation}
- \frac{1.5 \times 10^{-3}}{\pi} N(0) [\psi^{(2)}(y)]^2 \Gamma_1^s \biggl(\frac{v_{\rm F}}{2 \pi T} \frac{|\Delta|^2}{T} \biggr)^2 {\rm ln}\biggl(\frac{2a}{r_c} \biggr), 
\end{equation}
where $\Gamma_1 = F_{1s}/(1+F_{1s}/3)$ 
with the Landau parameter $F_{1s}$ ($\geq 0$). In obtaining eq.(15), the impurity scattering-induced vertex correction to the FL-corrected term was neglected because such a vertex correction results only in higher order terms in the gradient. Although this term, eq.(15), is of the same sign as that of eq.(14) and leads to an additional increase of the coefficient $K_1$, it is of a higher order in $|\Delta|^2$ and thus, is negligibly small in the close vicinity of the $T_c(P)$ curve, where $|\Delta|$ is the amplitude of the order parameter in the bulk liquid with no vortices. Therefore, as far as one focuses on the close vicinity of the $T_c(P)$-curve, this correlation-induced increase of the separation between the half cores may be neglected. 

\section{Summary}

In the present work, stability of the vortices in the B-phase of superfluid $^3$He in globally isotropic aerogel has been examined. First, it has been pointed out that the impurity-scatterings due to the aerogel structure should lead to a $|{\rm ln}T|$-growth of the coefficient of one type of the gradient term. This growth of one rigidity upon cooling does not affect the relative stability of the three types of vortices found in the bulk superfluid, while it significantly changes the characters of the two vortices to be experimentally realized. First, the spatial distribution of the core magnetization appearing in the axisymmetric v-vortex is qualitatively changed as the temperature is lowered along the $T_c(P)$-curve, i.e., as the pressure $P$ is lowered. However, it is unclear whether this result is observable because the strong-coupling corrections in the coefficients of the GL-quartic terms, which are necessary in describing events at higher pressures where the v-vortex really occurs, were neglected in the present work. 

As an event to be observable in the weak coupling approximation used here, we have pointed out that the impurity-scattering induced growth of one rigidity upon cooling enhances the anisotropy of the core structure of the double-core vortex to be realized at lower pressures. That is, the spacing between the half-cores grows as the pressure $P$ is lowered along the $T_c(P)$-line. Such a growth of the vortex core anisotropy has also been pointed out as a consequence of the Fermi liquid correction \cite{Th2}. In the case, 
as the half-cores are sufficiently separated from each other, and the Landau-Zener tunnelling between the two quasiparticle spectra near the double-core vortex center consequently increases, the spectrum of the low energy Fermi excitation around a double-core vortex changes to two localized ones at the half cores, and the resulting splitting of the quasiparticle spectra is reflected as the appearance of a slow mode of the rotational dynamics of a vortex. This mechanism changing the vortex dynamics is based largely on the increase of the separation between the half cores and hence, in the B phase in globally isotropic aerogels, is expected to be seen, according to the enhancement of the core anisotropy found in the present work, even with decreasing the pressure. In the case of the bulk liquid, application of a magnetic field perpendicular to the vortex axis tends to change the double-core vortex to the v-vortex \cite{Ohmi}. On the other hand, in isotropic aerogels, such a growth of the vortex core anisotropy upon cooling occurs in the weak coupling regime where the v-vortex is much higher in energy than the double-core vortex. Hence, application of a magnetic field in this case may not lead to realization of the v-vortex. In this manner, the two mechanisms of the growth of the core anisotropy in the double-core vortex may be distinguished experimentally in isotropic aerogels. 

\begin{acknowledgments}
The present research was supported by JSPS KAKENHI (Grant No.16K05444). 
\end{acknowledgments}

\end{document}